\documentclass[showkeys, aps, superscriptaddress, prmaterials, floatfix]{revtex4-2}

\usepackage{graphicx}%
\usepackage{amsmath,amssymb,amsfonts}%
\usepackage{amsthm}%
\usepackage{mathrsfs}%
\usepackage[dvipsnames]{xcolor}

\begin{document}

\title{Optimization of Epitaxial Mn$_4$N Thin Films Grown by Sputtering for Spintronic Applications}

\author{Teodor Apetrei}
\affiliation{Institut de Ciencia de Materials de Barcelona (ICMAB-CSIC), Campus de la UAB, 08193 Bellaterra, Barcelona, Spain}

\author{Emre Demiroglu}
\affiliation{Department of Physics, Marmara University, 34722, Ziverbey, Istanbul, Turkiye}

\author{Caner Deger}
\affiliation{Department of Physics, Marmara University, 34722, Ziverbey, Istanbul, Turkiye}

\author{Can Onur Avci}
\email{cavci@icmab.es}
\affiliation{Institut de Ciencia de Materials de Barcelona (ICMAB-CSIC), Campus de la UAB, 08193 Bellaterra, Barcelona, Spain}

\author{Silvia Damerio}
\email{sdamerio@icmab.es}
\affiliation{Institut de Ciencia de Materials de Barcelona (ICMAB-CSIC), Campus de la UAB, 08193 Bellaterra, Barcelona, Spain}

\keywords{spintronics, perpendicular magnetic anisotropy, ferrimagnets, spin-orbit torques, anomalous Hall effect, magnetoresistance}

\begin{abstract}
Ferrimagnetic Mn$_4$N has recently emerged as a promising rare-earth-free platform for spintronic devices, owing to its low magnetization, high domain wall mobility, and strong anomalous Hall response. However, achieving thin films with robust perpendicular magnetic anisotropy (PMA) and spin-orbit torque (SOT) functionality via scalable deposition methods remains a key challenge. Here, we systematically investigate the growth of Mn$_4$N thin films by reactive magnetron sputtering and establish the conditions required to obtain high-quality films with properties suitable for SOT applications. We show that epitaxial, single-crystalline Mn$_4$N with strong PMA can be achieved on MgO(100), while films grown on SrTiO$_3$(100) remain textured. The optimized films exhibit square-like hysteresis loops with high remanence, large and tunable coercivity, and a large anomalous Hall effect signal. By combining structural, magnetic, and magnetotransport measurements with density functional theory calculations, we identify epitaxial strain as an important tuning parameter for magnetic anisotropy, while also demonstrating that it is not the sole contributor. In particular, our results highlight the additional role of interfacial effects in stabilizing PMA. Finally, we demonstrate efficient current-induced magnetization switching in Mn$_4$N/Pt bilayers, confirming efficient interfacial spin transparency. These results establish sputtered Mn$_4$N as a viable and versatile material platform for energy-efficient spin-orbitronic devices.
\end{abstract}

\maketitle

\section{Introduction}
Spin-orbit torques (SOTs) provide an efficient route to manipulate magnetization electrically and are at the core of next-generation spintronic devices for memory and logic applications.~\cite{manchon2019,ryu2020current,shao2021} In this context, materials with perpendicular magnetic anisotropy (PMA) are desirable, as they enable high-density non-volatile memories,~\cite{Piramanayagam2007JAP} high-thermal-stability magnetic tunnel junctions,~\cite{Ikeda2010,Cubukcu2014APL} efficient current-driven domain-wall motion,~\cite{miron2011fast} and the stabilization of chiral spin textures such as Néel domain walls and skyrmions.~\cite{Emori2013,Finocchio_2016} Among the various material platforms, ferrimagnets have attracted significant attention due to their ability to sustain ultrafast dynamics and high domain-wall velocities, making them promising candidates for high-performance and energy-efficient devices.~\cite{kim2022ferrimagnetic}

Most studies to date have focused on rare-earth-based ferrimagnetic alloys and garnets, which exhibit excellent spintronic properties,~\cite{finley2020spintronics,emori2021ferrimagnetic,avci2021current,zhou2021efficient,zhang2023ferrimagnets} but, the majority suffer from material scarcity and integration challenges. In this regard, the transition-metal nitride Mn$_4$N has emerged as a compelling rare-earth-free alternative.~\cite{ito2021,Zhang_2022} This material combines a ferrimagnetic ground state with relatively low saturation magnetization, large anomalous Hall effect (AHE),~\cite{shen2014metallic} and efficient current-driven magnetization dynamics.~\cite{Gushi2019STT,Bai2022,ghosh2021current} In addition, its compatibility with thin-film growth techniques and the possibility to tune its magnetic properties via strain, composition, and interfaces make Mn$_4$N particularly attractive for SOT-based devices.~\cite{mitarai2020,ito2020,damerio2025magneto}

Despite these advantages, achieving Mn$_4$N thin films with robust PMA and reproducible spintronic functionality using scalable deposition techniques remains challenging. To date, molecular beam epitaxy (MBE) has been the most established approach to obtain high-quality epitaxial Mn$_4$N films with strong PMA and excellent transport properties.~\cite{yang2002crystalline,yasutomi2014perpendicular,ito2016,komori2019magnetic} In contrast, reactive magnetron sputtering -a more scalable and industrially compatible method- has been comparatively less explored. Initial studies on sputtering deposition of epitaxial Mn$_4$N have predominantly focused on relatively thick films (typically $\gtrsim$100 nm).~\cite{Kazuki2015} Extending sputtering growth to thinner films, in the range of a few tens of nanometers or below, which are required for spintronic devices and current-induced magnetization control, is non-trivial due to the increased sensitivity of structural and magnetic properties to growth conditions at reduced thickness. Only recently some reports on 30 nm~\cite{isogami2020contributions,xu2026orbital} and sub-20 nm~\cite{zhou2021rare} sputtered Mn$_4$N films have appeared, but a systematic understanding of how deposition parameters affect crystallinity, magnetic anisotropy, and especially magneto-transport and spintronic properties in this technologically relevant thickness regime is still lacking.

In this work, we address this challenge by systematically investigating the growth of Mn$_4$N thin films by reactive magnetron sputtering on MgO(100) and SrTiO$_3$(100) substrates. We correlate structural, magnetic, and magnetotransport properties across a wide range of deposition temperature and N$_2$ partial pressure during deposition, identifying the key parameters that enable the stabilization of high-quality films with strong PMA. We show that epitaxial, single-crystalline films can be grown on MgO, whereas growth on SrTiO$_3$ yields textured polycrystalline films with distinct magnetic behavior. By combining experiments with density functional theory calculations, we further demonstrate that, although epitaxial strain plays an important role in tuning magnetic anisotropy, it is not the sole mechanism, and interfacial effects must also be considered. Finally, we demonstrate efficient current-induced magnetization switching in Mn$_4$N/Pt bilayers, highlighting the potential of sputtered Mn$_4$N for spin-orbitronic applications.

\section{Methods}
Mn$_4$N thin films with thicknesses $t = 10$, 30, 50, and 100 nm were deposited by reactive magnetron sputtering on MgO(100) and SrTiO$_3$(100) substrates. The deposition was carried out at a DC power of 25 W, corresponding to a growth rate of 0.0167 nm/s, under a total reactive gas (Ar + N$_2$) pressure of 3 mTorr and with a target-substrate distance of 150 mm. The N$_2$ content during growth was controlled by systematically varying the Ar:N$_2$ gas ratio from 30:1 to 5:1.

To investigate the role of substrate temperature, depositions were performed in the range 400~$^\circ$C to 550~$^\circ$C, following prior reports on the optimal epitaxial Mn$_4$N growth by MBE around 450~$^\circ$C.~\cite{yasutomi2014perpendicular,Meng2015} To prevent surface oxidation, all films were capped \textit{in situ} with a 3 nm thick Pt or V layer.

The crystal structure was characterized by X-ray diffraction (XRD) using Cu K-$\alpha$ radiation ($\lambda = 1.5406$~\AA). Atomic-scale structural and compositional analyses were performed by high-resolution scanning transmission electron microscopy (STEM) combined with energy-dispersive X-ray spectroscopy (EDX), which allows elemental resolution. Magnetic properties were measured at room temperature using a superconducting quantum interference device (SQUID).

For magnetotransport measurements, Hall bar devices were patterned by optical lithography followed by ion milling, with channel dimensions of 30~$\mu$m~$\times$~7.5~$\mu$m and a Hall cross width of 3~$\mu$m. Longitudinal and transverse voltages were measured using a Zurich Instruments MFLI lock-in amplifier, with an AC current of 2 mA (r.m.s.) at 2.9 kHz supplied by a Keithley 6221 current source. Current-induced switching measurements were performed using 1 ms current pulses with varying amplitudes up to 45 mA, separated by 800 ms intervals, while applying an in-plane magnetic field parallel to the current direction to break the rotational symmetry of damping-like SOT, as customary in such measurements. The resistivity of the samples was calculated to be 2.9 $\mu\Omega$/cm, in good agreement with previously reported bulk values.~\cite{lu2016ternary}

First-principles calculations were performed within density functional theory using the Vienna ab initio simulation package (VASP).~\cite{kresse1993ab, kresse1996efficient} The exchange-correlation interaction was described within the generalized gradient approximation using the Perdew-Burke-Ernzerhof functional, while the electron-ion interaction was treated with the projector augmented-wave method.~\cite{kresse1999ultrasoft, perdew1996generalized, perdew2008restoring} A plane-wave cutoff energy of 400 eV was used for all calculations. The investigated system was ferrimagnetic Mn$_4$N, where the Mn-I and Mn-II sublattices were initialized with antiparallel magnetic moments according to the ferrimagnetic ground state. The 2 × 2 × 2 k-point mesh was found to provide converged total energies for the considered supercell. Structural relaxations were performed without spin-orbit coupling (SOC) until the residual forces were below 0.01 eV/Å, and the electronic self-consistency criterion was set to $10^{-6}$ eV in total energy. To investigate the effect of epitaxial strain on the magnetic anisotropy, tetragonally distorted Mn$_4$N  structures were generated by imposing different a/c ratios, including the strain states corresponding to Mn$_4$N  matched to STO and MgO substrates. SOC was then included, and two self-consistent calculations were performed by constraining the ferrimagnetic spin axis along the out-of-plane and in-plane directions. The magnetic anisotropy energy was obtained from the total-energy difference between the in-plane and out-of-plane magnetization configurations,~\cite{demiroglu2024oscillatory} normalized by the number of magnetic Mn atoms in the simulation cell. With this definition, a positive magnetic anisotropy energy indicates that the out-of-plane configuration is energetically favored, corresponding to perpendicular magnetic anisotropy.

\section{Results and discussion}
\subsection{X-ray characterization}

Mn$_4$N is a transition metal nitride with a cubic antiperovskite structure,~\cite{takei1960,takei1962} where nitrogen occupies the body-center position of a face-centered cubic Mn lattice (space group $Pm\overline{3}m$). Epitaxial growth on MgO and SrTiO$_3$ (STO) substrates is enabled by the relatively good lattice matching with rock-salt and perovskite structures, respectively. The nominal lattice mismatch corresponds to a tensile strain of approximately $-8\%$ for Mn$_4$N(100)/MgO(100) and $-1\%$ for Mn$_4$N(100)/STO(100).~\cite{yasutomi2014perpendicular, Kazuki2015} However, due to the richness of the Mn$_x$N$_y$ phase diagram, precise control over growth parameters such as substrate temperature and N$_2$ partial pressure is required to stabilize single-phase Mn$_4$N.~\cite{Zhang_2022}

\begin{figure}[htb]
  \centering
    \includegraphics[width=1\textwidth]{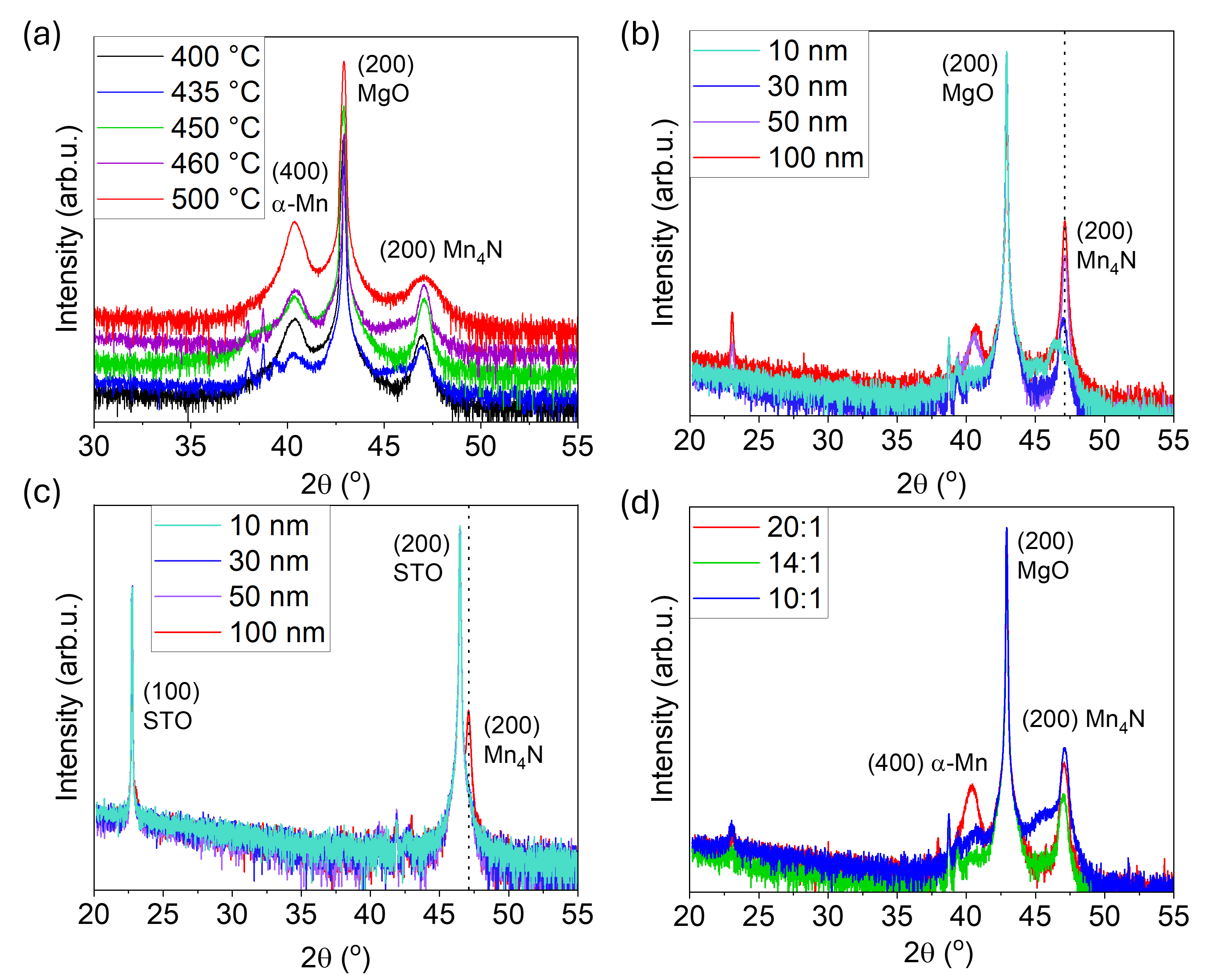}
    \caption{XRD spectra of Mn$_4$N films. $\theta-2\theta$ scan of film grown at (a) different deposition temperature on MgO; (b) different thicknesses on MgO substrate; (c) different thicknesses on STO substrate; (d) different deposition Ar:N${_2}$ gas ratios on MgO substrate. The dotted lines indicate the (200) peak position in bulk. All samples are capped with Pt.}
    \label{fig:XRD}
\end{figure}

X-ray diffraction (XRD) measurements were performed in the $2\theta$ range 20$^\circ$--80$^\circ$, focusing on 20$^\circ$--55$^\circ$, where the Mn$_4$N reflections are expected. Figure~\ref{fig:XRD}(a) shows the XRD patterns of 30 nm thick films grown on MgO(001) at different substrate temperatures. A dominant peak at $2\theta \approx 47^\circ$, corresponding to the Mn$_4$N (200) reflection, is observed, indicating preferential (100) orientation of the films. A second peak at $2\theta \approx 40.3^\circ$, attributed to the (400) reflection of $\alpha$-Mn, \cite{isogami2020contributions} reveals the presence of a secondary phase, indicating incomplete nitridation during growth. As the temperature increases, the Mn$_4$N peak intensity maximizes while the $\alpha$-Mn contribution is suppressed, pointing to an optimal growth temperature around 450~$^\circ$C.

The evolution of the XRD patterns with film thickness on MgO is shown in Fig.~\ref{fig:XRD}(b). With decreasing thickness, the Mn$_4$N (200) peak shifts towards lower angles, corresponding to an increase in the out-of-plane lattice parameter from 3.85~\AA{} (100 nm) to 3.9~\AA{} (10 nm), compared to the bulk value of 3.86.~\AA{}~\cite{ito2021} This trend is consistent with the presence of epitaxial strain, which is more pronounced in thinner films (up to $\sim$1\%) and gradually relaxes with increasing thickness (to $\sim$0.25\% at 100 nm). Such strain levels are consistent with previous reports and are known to contribute to the emergence of perpendicular magnetic anisotropy (PMA) in Mn$_4$N.~\cite{yasutomi2014perpendicular,hirose2020strong} In addition, the $\alpha$-Mn phase becomes detectable only for thicknesses above 30 nm.

For films grown on STO substrates [Fig.~\ref{fig:XRD}(c)], the Mn$_4$N (200) reflection overlaps with the substrate peak at $2\theta = 46.49^\circ$, complicating the analysis for thinner films. For thicknesses above 50 nm, a distinct peak emerges at $2\theta \approx 47.08^\circ$, indicating the development of a (200)-textured Mn$_4$N phase. Overall, XRD characterization on STO is less conclusive at low thicknesses compared to MgO, where the film peaks are well separated from the substrate contribution.

Finally, the effect of the nitrogen content during growth is shown in Fig.~\ref{fig:XRD}(d) for films deposited on MgO. At low nitrogen content (e.g., Ar:N$_2$ = 20:1 and below), the formation of the secondary $\alpha$-Mn phase is favored, indicating insufficient nitridation. Increasing the N$_2$ partial pressure to moderate values (e.g., 14:1) promotes the formation of Mn$_4$N, while at higher N$_2$ content (10:1 or higher) a broad feature appears around $2\theta \approx 45^\circ$, suggesting the possible formation of other manganese nitride phases. From this analysis, we identify a substrate temperature of $\sim$450~$^\circ$C and an Ar:N$_2$ ratio of 14:1 as optimal growth conditions  Mn$_4$N.

\subsection{STEM characterization}

To further assess the role of the substrate in the crystal structure at the atomic scale, we performed high-resolution STEM on representative 30 nm-thick films grown on MgO and STO at optimum conditions.

\begin{figure}[htb]
  \centering
    \includegraphics[width=1\textwidth]{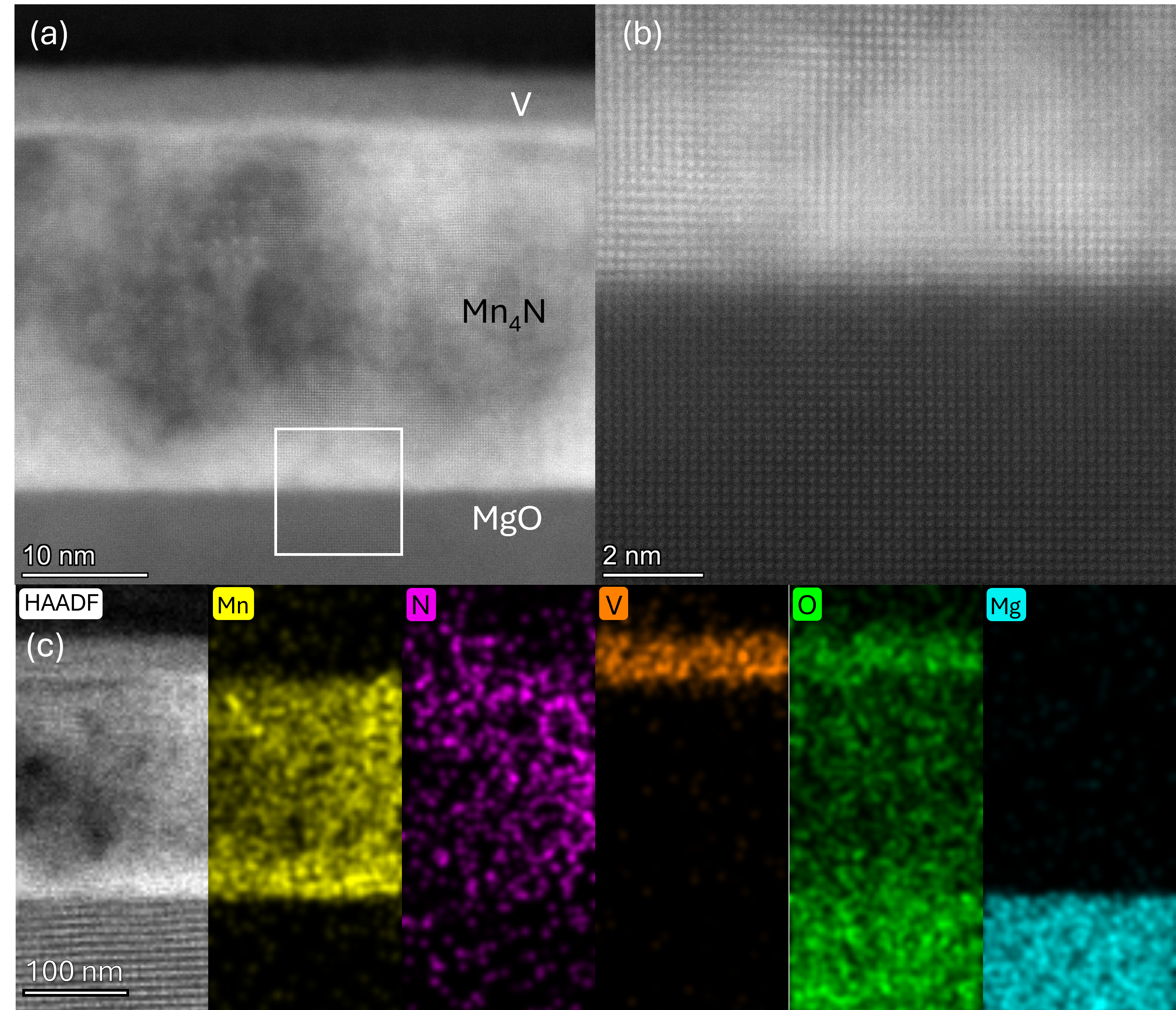}
    \caption{STEM characterization of a 30 nm thick Mn$_4$N film on MgO substrate with V capping. (a) Overview STEM image of a cross-section of a 30 nm Mn$_4$N film. (b) Enlarged STEM image of the interface of the Mn$_4$N film with the MgO substrate corresponding to the area within the white square in the overview image. (c) Elemental maps of the film composition measured via EDX.}
    \label{fig:TEM}
\end{figure}

For the film grown on MgO, the STEM images [Fig.~\ref{fig:TEM}] reveal a highly ordered, continuous lattice extending throughout the entire film thickness, consistent with epitaxial growth. The square atomic arrangement characteristic of the Mn$_4$N structure is clearly visible in the overview image [Fig.~\ref{fig:TEM}(a)]. A sharp interface with the MgO substrate is observed [Fig.~\ref{fig:TEM}(b)], with a well-defined stacking relationship in which 16 Mn atomic planes accommodate 17 Mg planes, indicative of periodic misfit dislocations that likely facilitate strain relaxation. Consistent with this picture, both in-plane and out-of-plane lattice parameters extracted from STEM ($3.8 \pm 0.1$~\AA{}) are close to the bulk value, indicating that the film is largely relaxed at this thickness.

Elemental mapping by EDX [Fig.~\ref{fig:TEM}(c)] shows a clear spatial separation between Mn and Mg, confirming limited intermixing at the interface. A partial interdiffusion of the V capping layer into the top $\sim$2 nm of the Mn$_4$N film is observed. While nitrogen is detected in the same regions as Mn, the overlap between N and O signals, combined with the air exposure during lamella preparation, results in a non-negligible oxygen background across the TEM specimen. This signal does not reflect the intrinsic chemical state of the as-grown Mn$_4$N thin films. Instead, it highlights the sensitivity of Mn$_4$N to oxidation and the importance of protective capping layers. Furthermore, we observe a minor degree of nitrogen redistribution in the V capping layer, possibly due to its strong N-affinity, which could give rise to N deficiencies in the top layers of the Mn$_4$N film. However, if present, such deficiencies do not affect either the macroscopic or local magnetic properties of the samples. Overall, these observations confirm the high crystalline quality and epitaxial nature of Mn$_4$N films grown on MgO, with only minor spatial variations in density across the thickness.

\begin{figure}[htb]
  \centering
    \includegraphics[width=1\textwidth]{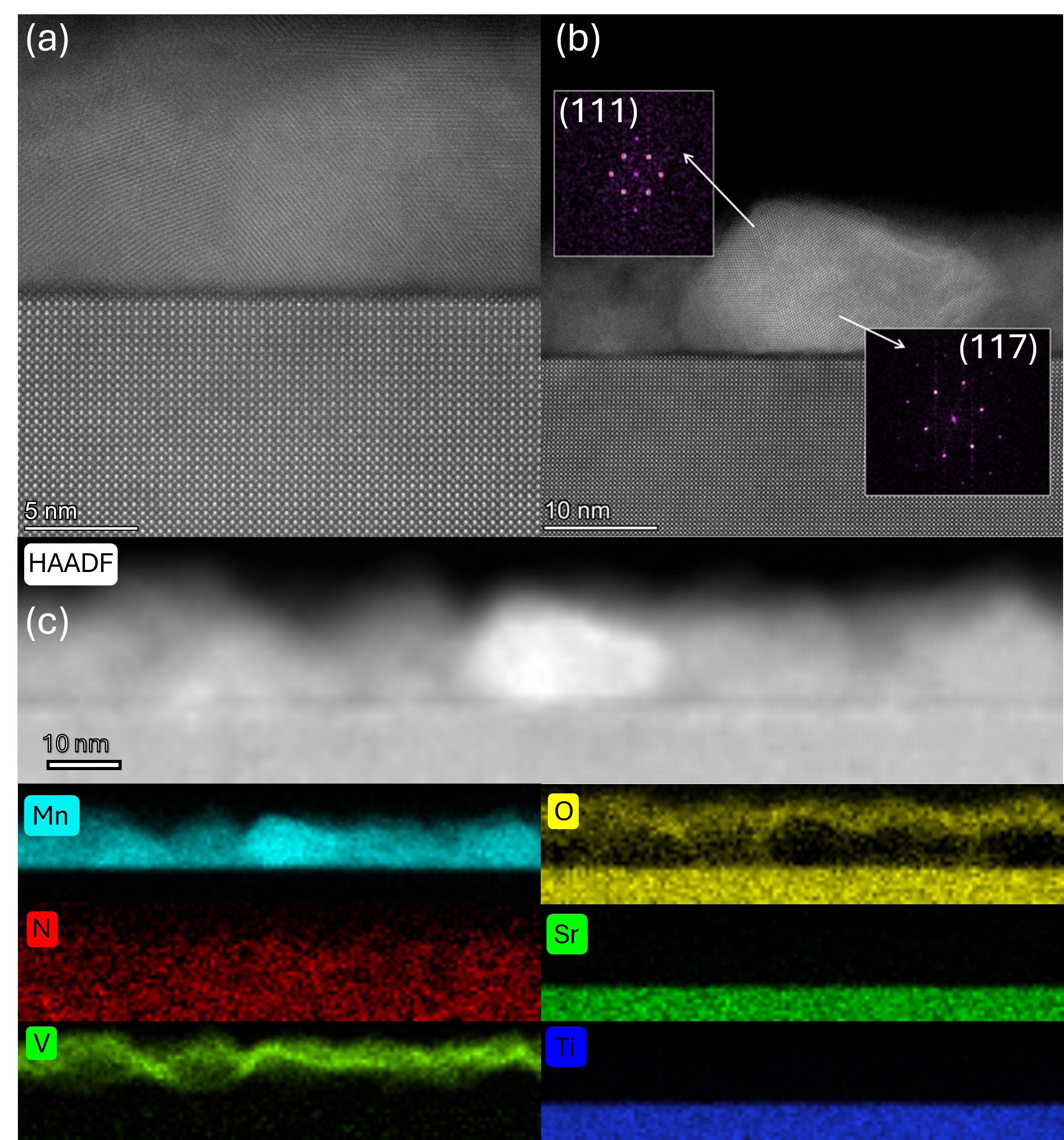}
    \caption{STEM characterization of a 30 nm thick Mn$_4$N film on STO substrate with V capping. (a) Overview STEM image of a cross section of a 30 nm Mn$_4$N film. (b) Enlarged STEM image of the polycrystalline grainy nature of Mn$_4$N and the STO substrate. The insets show the fast fourier transform (FFT) of two grains with (111) and (117) out-od-plane orientation. (c) Elemental maps of the film composition measured via EDX. }
    \label{fig:TEMSTO}
\end{figure}

In contrast, films grown on STO exhibit a markedly different morphology. As shown in Fig.~\ref{fig:TEMSTO}(a), the Mn$_4$N layer consists of multiple grains with distinct crystallographic orientations, indicating textured polycrystalline growth rather than epitaxy. A region of reduced contrast is observed near the interface, which may indicate a defective layer, possibly associated with vacancies, that could hinder coherent nucleation. Fast Fourier transform (FFT) analysis [Fig.~\ref{fig:TEMSTO}(b)] reveals a distribution of out-of-plane orientations, including (100), (111), (011), and (117), all with lattice parameters consistent with relaxed Mn$_4$N ($3.8 \pm 0.1$~\AA{}). EDX analysis [Fig.~\ref{fig:TEMSTO}(c)] shows no significant compositional differences between grains of different orientations, confirming that they belong to the same Mn$_4$N phase.

The emergence of multiple grain variants on STO, despite its smaller nominal lattice mismatch with Mn$_4$N compared to MgO, highlights that factors beyond simple lattice matching—such as interface quality and growth kinetics—play a decisive role in determining the film microstructure.

\subsection{Magnetic properties}

\begin{figure}[htb]
  \centering
    \includegraphics[width=1\textwidth]{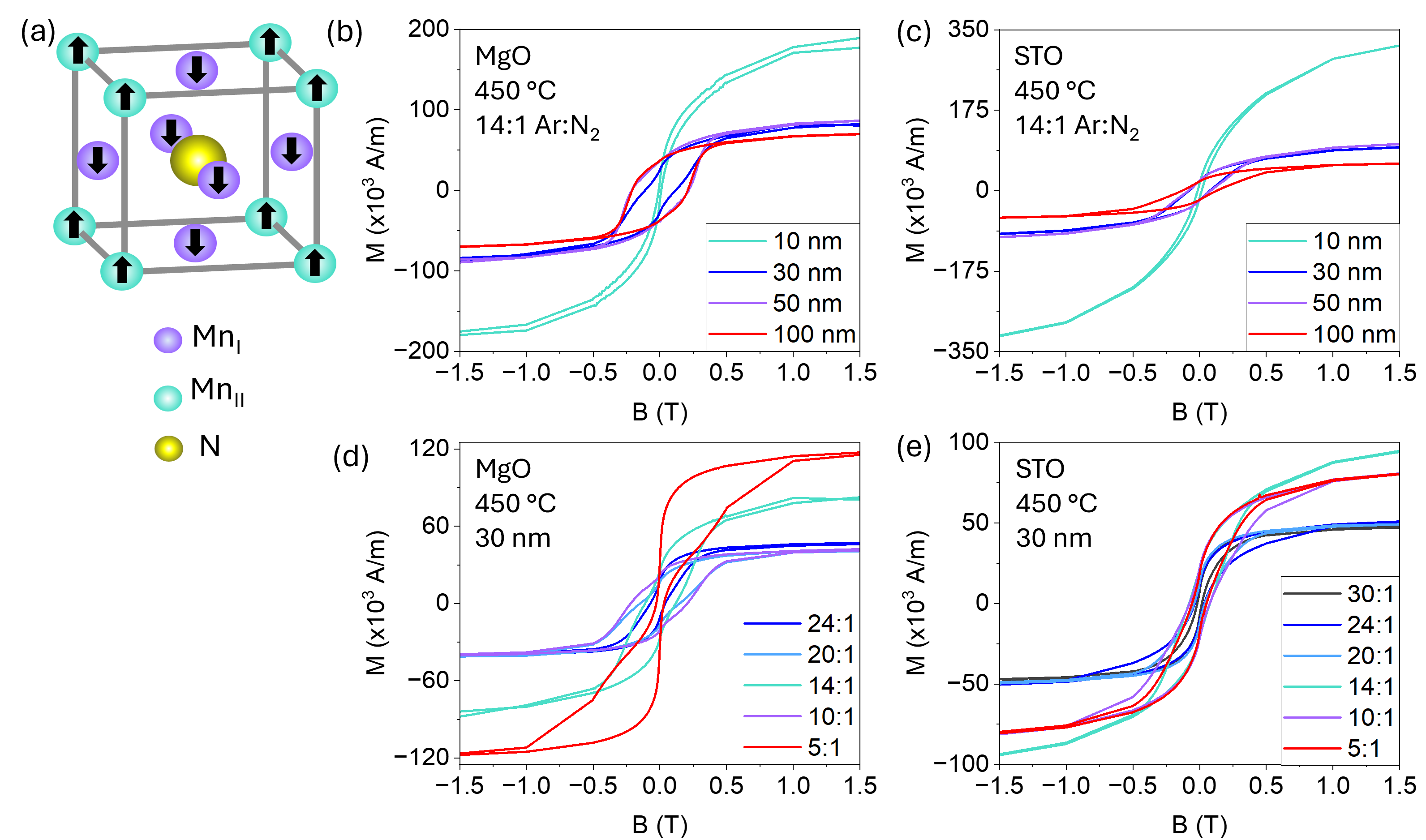}
    \caption{Ferrimagnetic properties. (a) Schematic representation of the Mn$_4$N unit cell, showing the two antiparallel Mn sublattices MnI (blue) and MnII (gray) and N atoms (red). Adapted from Ref.\cite{shen2014metallic}. The electron spin orientation is indicated by the black arrows.  Magnetization ($M$) vs out-of-plane field ($B_Z$) loops of 30 nm thick Mn$_4$N films grown at 450 ºC and 14:1 Ar:N$_2$ with varied thicknesses on (b) MgO and (c) STO substrate. $M-B$ loops of 30 nm thick Mn$_4$N films grown at 450 ºC  with varied deposition gas ratios on (d) MgO and (e) STO substrate. All samples are capped with Pt.}
    \label{fig:squid}
\end{figure}

Bulk Mn$_4$N is a ferrimagnet with a Curie temperature of approximately 740 K. Its magnetic structure arises from antiferromagnetic coupling between Mn atoms at the cube corners and those at the face centers of the unit cell [Fig.~\ref{fig:squid}(a)], resulting in a net room-temperature magnetization of $\sim 1.45 \times 10^5$ A/m.~\cite{Zhang_2022} When grown as thin films, Mn$_4$N can exhibit perpendicular magnetic anisotropy (PMA),~\cite{Li_2022,yasutomi2014perpendicular} although the magnetic properties are highly sensitive to deposition conditions, substrate choice, and thickness. Understanding these dependencies is essential for spin-orbitronic applications.

Figures~\ref{fig:squid}(b,c) show out-of-plane SQUID measurements of Mn$_4$N films with varying thickness grown on MgO and STO substrates, respectively. For thicknesses above $\sim$20 nm, films on both substrates exhibit clear hysteresis with high remanence at zero field with ratios $\sim$80\% and coercivity around 0.25 T, consistent with previous reports.~\cite{yasutomi2014perpendicular} The saturation magnetization is $\sim 8.5 \times 10^4$ A/m and decreases slightly with increasing thickness. The reduced saturation magnetization $M_S$ of materials in thin film form compared to bulk is not uncommon and is usually attributed to a combination of finite-size effects and interfacial magnetic dead layers, which predominates at lower thickness. In contrast, the reduced remanence and enhanced $M_S$ observed in the 10 nm-thick films point to a different mechanism underlying the deviation from the bulk magnetic properties (see Supplementary Information section 1).
Notably, strong PMA persists even in thicker films ($>30$ nm), where the lattice is largely relaxed, indicating that epitaxial strain is not the sole origin of magnetic anisotropy in this system. This observation points to additional contributions, including interfacial effects which are enhanced when the adjacent layer possessing strong SOC.~\cite{damerio2025magneto,lu2016ternary}

The influence of N$_2$ content during growth is summarized in Fig.~\ref{fig:squid}(d,e). For films on MgO [Fig.~\ref{fig:squid}(d)], robust PMA is maintained over a broad range of Ar:N$_2$ ratios (10:1 to 24:1), with remanence values of 75\%-90\% and coercivities between 0.1 and 0.3 T. At high N$_2$ content (5:1), the hysteresis loops become pinched, indicating multidomain formation and degraded magnetic uniformity, accompanied by an increase in $M_S$. In contrast, films grown on STO [Fig.~\ref{fig:squid}(e)] exhibit a weaker dependence on N$_2$ ratio, consistent with their polycrystalline nature. In this case, the coercivity increases with increasing N$_2$ content, while $M_S$ is higher for films grown under moderate Ar:N$_2$ conditions.

\begin{figure}[htb]
  \centering
    \includegraphics[width=1\textwidth]{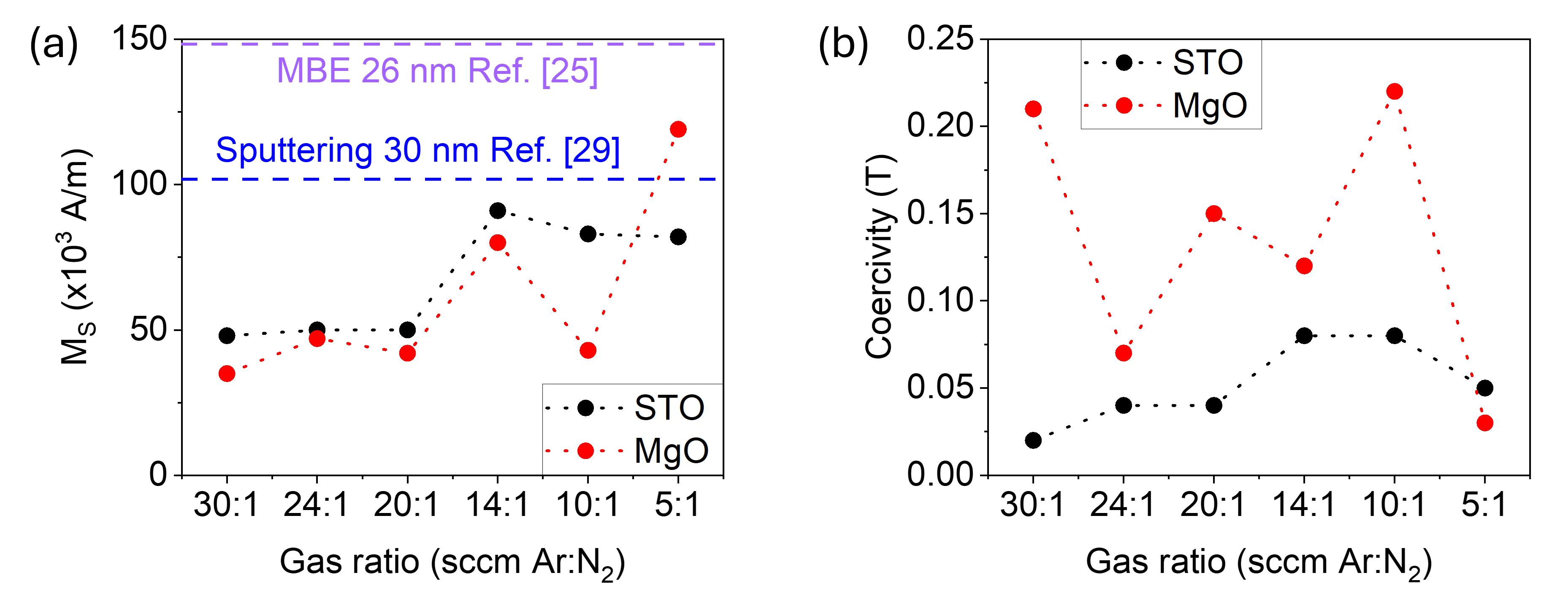}
    \caption{ Summary of the dependence of (a) saturation magnetization ($M_S$) and (b) coercivity on the gas ratio. All samples are capped with Pt.}
    \label{fig:squid2}
\end{figure}

A summary of the evolution of $M_S$ and $H_C$ with growth parameters is presented in Fig.~\ref{fig:squid2}(a,b). For comparable deposition conditions, similar values of $M_S$ are obtained on both substrates. Optimized films (Ar:N$_2$ = 14:1 30 nm) exhibit $M_S$ values of $\sim 8.0 \times 10^4$ A/m (MgO) and $\sim 9.1 \times 10^4$ A/m (STO), in reasonable agreement with previously reported values for sputtered Mn$_4$N,~\cite{Kazuki2015,isogami2020contributions} although still lower than those achieved by MBE~\cite{yasutomi2014perpendicular} or PLD.~\cite{shen2014metallic} In contrast, the coercivity shows a strong dependence on growth conditions for MgO, but remains comparatively insensitive when grown on STO. This highlights the higher sensitivity of epitaxial films to deposition parameters and underscores the importance of precise growth control for optimizing magnetic properties.

\subsection{Density Functional Theory}
These observations suggest that, although epitaxial strain contributes to the magnetic anisotropy, it cannot fully account for the experimentally observed PMA. While Isogami and coworkers~\cite{isogami2020contributions} demonstrated how nitrogen content also contributes to PMA, this is not the case of the films of the present study, grown in the same Ar:N$_2$ ratio but on different substrates. Therefore, to gain further insight into the role of strain and its effect on the magnetic anisotropy energy, we perform density functional theory calculations.

\begin{figure}[htb]
  \centering
    \includegraphics[width=1\textwidth]{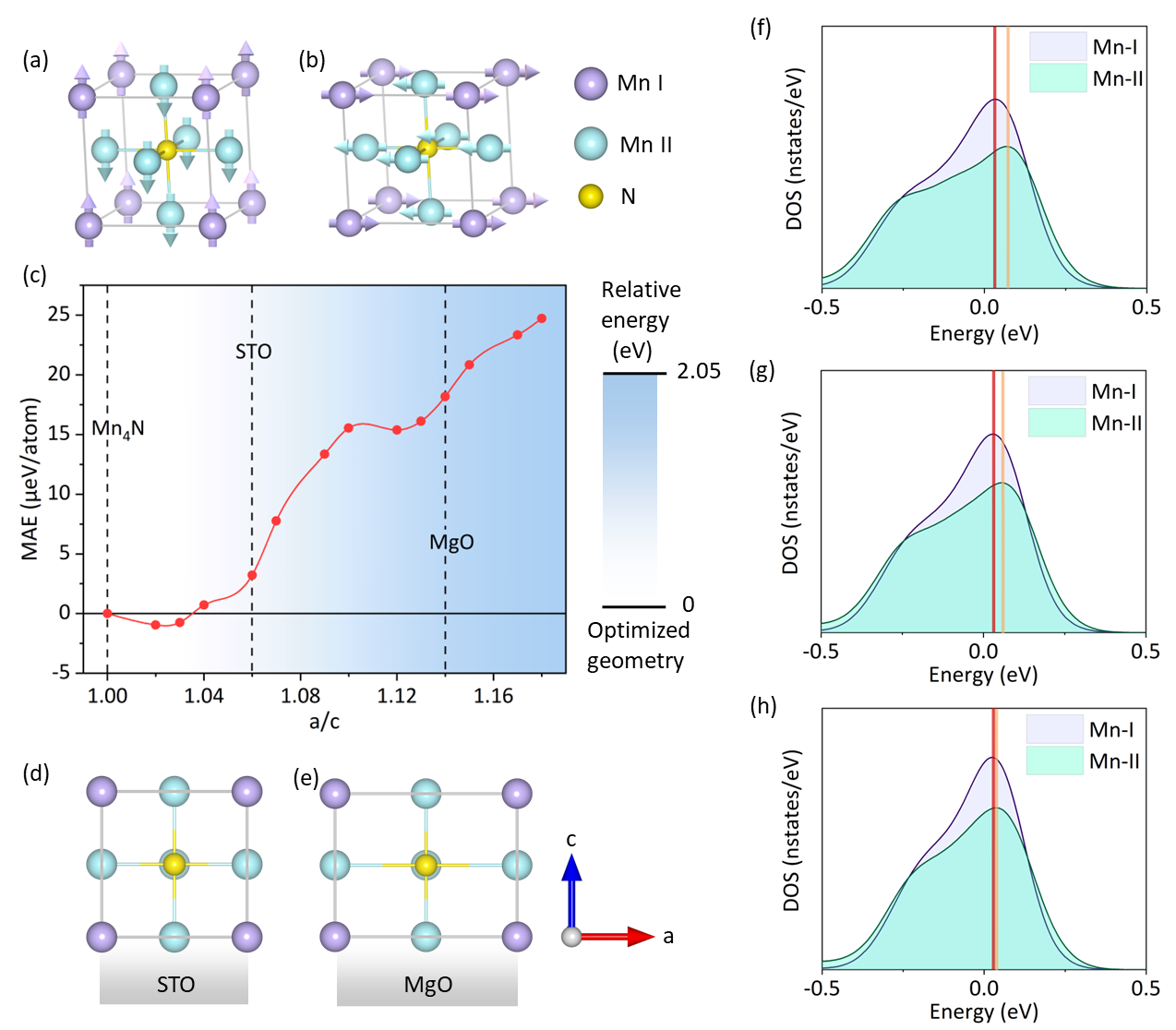}
    \caption{(a,b) Spin configurations used for magnetic anisotropy calculations in Mn$_4$N: (a) out-of-plane (OOP) and (b) in-plane (IP) magnetization orientations. (c) Magnetic anisotropy energy (MAE) as a function of tetragonal distortion (a/c) for bulk Mn$_4$N under strain conditions mimicking epitaxial growth on STO and MgO substrates. The background color gradient represents the relative elastic energy cost per unit cell required to reach the corresponding strain state. (d,e) Schematic representations of Mn$_4$N epitaxially strained to match (d) STO and (e) MgO lattice parameters. (f–h) Atom-resolved density of states (DOS) near the Fermi level: (f) optimized bulk Mn$_4$N, (g) strain state corresponding to STO, and (h) strain state corresponding to MgO. In all cases, noticeable Mn-I and Mn-II hybridization is observed around the Fermi level.}
    \label{fig:DFT}
\end{figure}

The influence of epitaxial strain on the magnetic anisotropy of Mn$_4$N  was examined by calculating the magnetic anisotropy energy (MAE) from the total-energy difference between ferrimagnetic configurations with out-of-plane and in-plane spin orientations, as illustrated in Fig.~\ref{fig:DFT}(a,b). The resulting MAE as a function of tetragonal distortion, expressed by the a/c ratio, is shown in Fig.~\ref{fig:DFT}(c), where the vertical dashed lines indicate the optimized Mn$_4$N  geometry and the strain states corresponding to epitaxial matching with STO and MgO substrates. Around the equilibrium structure, Mn$_4$N  exhibits only weak magnetic anisotropy, with MAE values close to zero; however, increasing the a/c ratio drives the MAE to positive and progressively larger values, demonstrating that tensile tetragonal distortion stabilizes an out-of-plane easy axis and promotes perpendicular magnetic anisotropy. The STO-like strain state yields a moderate positive MAE, whereas the MgO-like distortion produces a substantially larger MAE, consistent with the experimentally observed larger anisotropy field in Mn$_4$N  films grown on MgO compared with those grown on STO. The background color map in Fig.~\ref{fig:DFT}(c), representing the relative elastic energy cost of each imposed distortion, shows that although larger a/c ratios require a higher elastic penalty, the substrate-relevant strain states remain energetically accessible. The schematic structures in Fig.~\ref{fig:DFT}(d,e) illustrate that epitaxial matching modifies the Mn-N bond lengths and Mn-Mn distances, thereby altering the local crystal-field environment and orbital overlap between the antiparallel Mn sublattices. This microscopic picture is further supported by the atom-resolved DOS shown in Fig.~\ref{fig:DFT}(f-h), which reveals how the electronic structure near the Fermi level evolves under substrate-induced tetragonal distortion.~\cite{aksu2020strain} For the relaxed Mn$_4$N  structure, Fig.~\ref{fig:DFT}(f) shows that Mn-I and Mn-II states are both present near the Fermi level, but their overlap remains relatively weak, consistent with the small MAE obtained close to the equilibrium geometry. When Mn$_4$N  is strained to emulate epitaxial growth on STO, as show in Fig.~\ref{fig:DFT}(g), the Mn-I and Mn-II contributions become more strongly hybridized around the Fermi level compared with the relaxed structure. This enhanced Mn-I/Mn-II overlap is consistent with the moderate increase in PMA observed for the STO-like strain state. In the MgO-like strained configuration, Fig.~\ref{fig:DFT}(h), the Mn-I and Mn-II spectral features show a much stronger overlap near the Fermi level, indicating a pronounced strain-induced modification of the electronic states that govern magnetic anisotropy. Since SOC-driven anisotropy is highly sensitive to the orbital character and hybridization of states close to the Fermi level, the stronger Mn-I/Mn-II hybridization in the MgO-like structure provides a microscopic explanation for the substantially enhanced positive MAE and the stabilization of perpendicular magnetic anisotropy. Although a quantitative discrepancy remains between the DFT-calculated MAE ratio and the experimentally extracted anisotropy-field ratio, likely due to finite-size/surface effects (see Supplementary Information section 2), strain relaxation, and film microstructure, both approaches show the same qualitative trend: MgO-like strain yields a larger perpendicular magnetic anisotropy than STO-like strain.

\subsection{Magnetotransport and spintronics properties}
\begin{figure}[htb]
  \centering
    \includegraphics[width=1\textwidth]{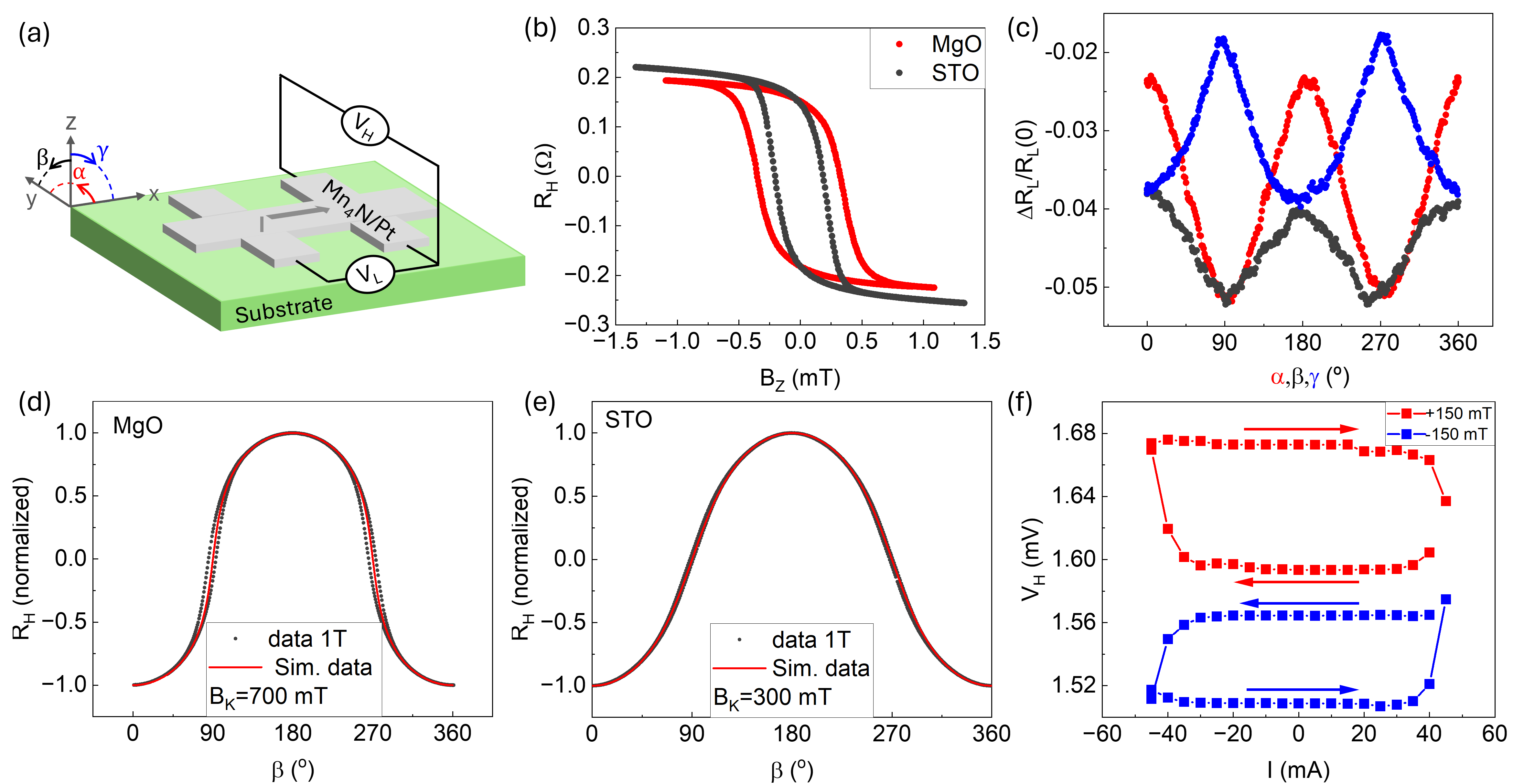}
    \caption{Magnetotransport properties of  Mn$_4$N(30)/Pt(3) bilayers. (a) Micrograph of the device and scheme of the connections; (b) Anomalous Hall resistance ($R_H$) as a function of out-of-plane magnetic field ($B_Z$) on MgO (black) and STO (red). (c) Longitudinal resistance ($R_L$) measured as a function of the direction of an external 1.4 T field, rotated in the xy ($\alpha$), zy ($\beta$) and zx ($\gamma$) planes. (d,e) $R_H$ as a function of $\beta$ measured in 1 T field for samples on MgO and STO, respectively. (f) Plot of the Hall voltage ($V_H$) as a function of applied current ($I$) with positive (red) and negative (blue) in-plane fields.}
    \label{fig:AMR}
\end{figure}

As the main goal of this study is to optimize the properties of sputtered Mn$_4$N thin films for spintronic applications, we investigate their magnetotransport behavior by fabricating Hall bar devices (30~$\mu$m $\times$ 7.5~$\mu$m), as shown in Fig.~\ref{fig:AMR}(a). These devices enable simultaneous measurements of the anomalous Hall effect (AHE) and anisotropic magnetoresistance (AMR), providing electrical access to the magnetic properties of the films.

Figure~\ref{fig:AMR}(b) shows representative AHE loops for 30 nm thick Mn$_4$N films grown on MgO and STO under optimized conditions. The hysteresis loops exhibit a progressive, slanted reversal over a broad field range, indicative of a magnetization reversal mechanism dominated by domain nucleation rather than domain-wall propagation. This interpretation is further supported by the remanence ratio of $\sim$93\%, which suggests the presence of residual multidomain states at zero applied field. The shape of the AHE loops closely matches that obtained from SQUID measurements, indicating that the reversal behavior observed in the Hall bar devices is representative of the magnetic response of the full films. Notably, the AHE amplitude reaches $\sim$40~m$\Omega$, exceeding previously reported values for sputtered Mn$_4$N,~\cite{Meng2015} indicating good electronic and magnetic quality.

The anisotropic magnetoresistance (AMR) of Mn$_4$N films grown on MgO is shown in Fig.~\ref{fig:AMR}(c). The longitudinal resistance $R_L$ follows the expected $\sin(\alpha)$ dependence for in-plane field rotation. In contrast, out-of-plane rotation leads to a non-sinusoidal response with pronounced cusps, reflecting the inability of the magnetization to fully align with the external field due to strong PMA. In addition to this behavior, a finite magnetoresistance is observed during the $\beta$ rotation, which can be attributed to spin Hall magnetoresistance arising from the adjacent Pt capping layer.~\cite{kim2016spin,xu2026orbital} This contribution likely originates from spin current reflection at the interface, with a possible enhancement due to interfacial hybridization and SOC.~\cite{kobs2011anisotropic,avci2019effects}

To quantitatively determine the anisotropy field ($B_K=\mu_0H_K$), we measured the angular dependence of the Hall resistance $R_H$ under a 1 T magnetic field for films on MgO and STO [Figs.~\ref{fig:AMR}(d,e)]. In both cases, deviations from a simple $\cos(\beta)$ behavior and the presence of hysteresis confirm a large perpendicular anisotropy. By fitting the data using micromagnetic simulations (details in the Supplementary Information section 3 and Ref.~\cite{testa2023leveraging}), we extract $B_K$$\approx 700$ mT for MgO and $B_K \approx 300$ mT for STO. This difference in $B_K$ is consistent with the difference of in-plane saturation field measured for films grown on the two substrates (see Supplementary Information section 4) and confirms stronger effective anisotropy for films grown on MgO. This also qualitatively agrees with the strain-dependent magnetic anisotropy energy obtained from DFT calculations discussed below. Moreover, the larger $B_K$ observed in MgO/Mn$_4$N/Pt compared to MgO/Mn$_4$N/V~\cite{damerio2025magneto} further supports a contribution of interfacial SOC to the overall magnetic anisotropy.

Finally, we demonstrate deterministic current-induced magnetization switching in Mn$_4$N(30)/Pt(3) bilayers grown on MgO. Notably, the ability to switch a relatively thick (30 nm) ferrimagnetic layer using interfacial SOTs highlights the efficient torque transfer at the Pt/Mn$_4$N interface. This thickness is comparable to state-of-the-art demonstrations of SOT switching in thick magnetic systems,~\cite{roschewsky2017spin,chen2024spin} underscoring the potential of sputtered Mn$_4$N for device-relevant implementations and high thermal energy barrier when combined with its large coercivity.

The magnetization state was probed via the Hall voltage $V_H$ following the application of 1 ms current pulses. As shown in Fig.~\ref{fig:AMR}(f), clear hysteretic switching is observed, with polarity reversal upon changing the direction of an in-plane magnetic field ($\pm$150 mT) applied parallel to the current, consistent with the symmetry of damping-like SOTs.~\cite{manchon2019,garello2013symmetry} Magnetization reversal occurs at a current of $\sim$40 mA, corresponding to a critical current density of $j_C \approx 1.8 \times 10^{7}$ A/cm$^2$, in line with our previous results on Mn$_4$N-based bilayers.~\cite{damerio2025magneto}

Importantly, the switching polarity is opposite to that observed in Mn$_4$N/Ta bilayers grown in similar conditions in our group,~\cite{damerio2025magneto} reflecting the opposite sign of the spin Hall angle in Pt and Ta.~\cite{Avci2014} Finally, the reduced resistance contrast observed in current-induced switching compared to field-driven loops is attributed to partial switching combined with Joule heating during the current pulses, the latter leading to a transient increase in temperature and consequently a reduction of the AHE amplitude and coercivity (details in the Supplementary Information section 5). 

\section{Conclusions}

In summary, we have established a reliable pathway to grow Mn$_4$N thin films by reactive magnetron sputtering with properties suitable for spintronic applications. We identify key growth parameters that enable high-quality films and show that epitaxial Mn$_4$N with strong perpendicular magnetic anisotropy (PMA) can be grown on MgO substrates, whereas growth on SrTiO$_3$ yields textured polycrystalline films.

By combining structural, magnetic, and transport measurements, we demonstrate that PMA in sputtered Mn$_4$N is robust and persists beyond the fully strained regime. Together with density functional theory calculations, our results indicate that while epitaxial strain contributes to magnetic anisotropy, interfacial effects also play a crucial role. Furthermore, we demonstrate efficient current-induced magnetization switching in 30 nm thick Mn$_4$N/Pt bilayers, highlighting the potential of this system for spin--orbit torque applications.

These results open several avenues for future work. In particular, further reducing the film thickness while maintaining strong PMA will be essential for device integration. Combining improved crystalline quality (achievable for example by employing facing-target sputtering configurations~\cite{nakagawa1994preparation, zhang2022microstructure}) with compositional tuning (such as by Ni incorporation~\cite{ghosh2021current}) will be essential to access a regime of controlled domain-wall propagation. In addition, engineering interfacial spin-orbit interactions to induce Dzyaloshinskii-Moriya interaction may allow stabilization of Néel-type domain walls and skyrmions in ultrathin Mn$_4$N-based heterostructures.~\cite{ZhouDMI} Overall, our work establishes sputtered Mn$_4$N as a promising and experimentally accessible platform for next-generation spintronic and spin-orbitronic devices.

%

\section*{Author contributions}
C.O.A. and S.D. conceived and supervised the study. T.A. performed the experiments, carried out the data analysis, and wrote the original draft of the manuscript. S.D. contributed to the experimental design, data analysis, and interpretation. E.D. and C.D. performed the density functional theory calculations. C.O.A. and S.D. revised the manuscript. All authors discussed the results and approved the final version of the manuscript.

\section*{Acknowledgments}
C.O.A. and S.D. acknowledge funding from the European Research Council (ERC) under the European Union’s Horizon 2020 research and innovation programme (project MAGNEPIC, Grant Agreement No. 949052). T.A., C.O.A. and S.D. acknowledge funding from the MUST project with reference PCI2023-143400/MCIU/AEI/10.13039/501100011033 co-funded by the European Union and the MAI-SKY project with reference PID2021-125973OA-I00 funded by MCIN/AEI/ 10.13039/501100011033/FEDER, UE. S.D. acknowledges funding from the Marie Sklodowska-Curie Actions (MSCA) under the European Union’s Horizon Europe research and innovation programme (project SPINDY, Grant Agreement No. 101106885). E.D. and C.D. acknowledge support from the Scientific and Technological Research Council of Türkiye (TÜBİTAK) under Grant No. 123N420 and from the Marmara University Scientific Research Projects Coordination Unit (BAP) under Project No. FBA-2025-11656. We acknowledge financial support from the State Investigation Agency, through the Severo Ochoa Programme for the Centers of Excellence in R$\&$D (CEX2023-001263-S). We acknowledge the Joint Electron Microscopy Center at ALBA (JEMCA) and funding from Grant IU16-014206 (METCAM-FIB) to ICN2 from the European Union through the European Regional Development Fund (ERDF), with the support of the Ministry of Research and Universities, Generalitat de Catalunya. We also acknowledge the ICMAB-CSIC Scientific and Technical Services: STEM (S. Estandia); X-ray Diffraction Laboratory (F. J. Campos Lopez) and Low Temperatures and Magnetometry (B. Bozzo Closas and F. Valles Perez).

\section*{Competing interests}
All authors declare no financial or non-financial competing interests. 

\section*{Data availability}
The data that support the findings of this study are available from the corresponding authors upon reasonable request.

\bibliography{Bibliography}  
\clearpage
\begin{center}
\includegraphics[page=1,width=\textwidth,height=\textheight,keepaspectratio]{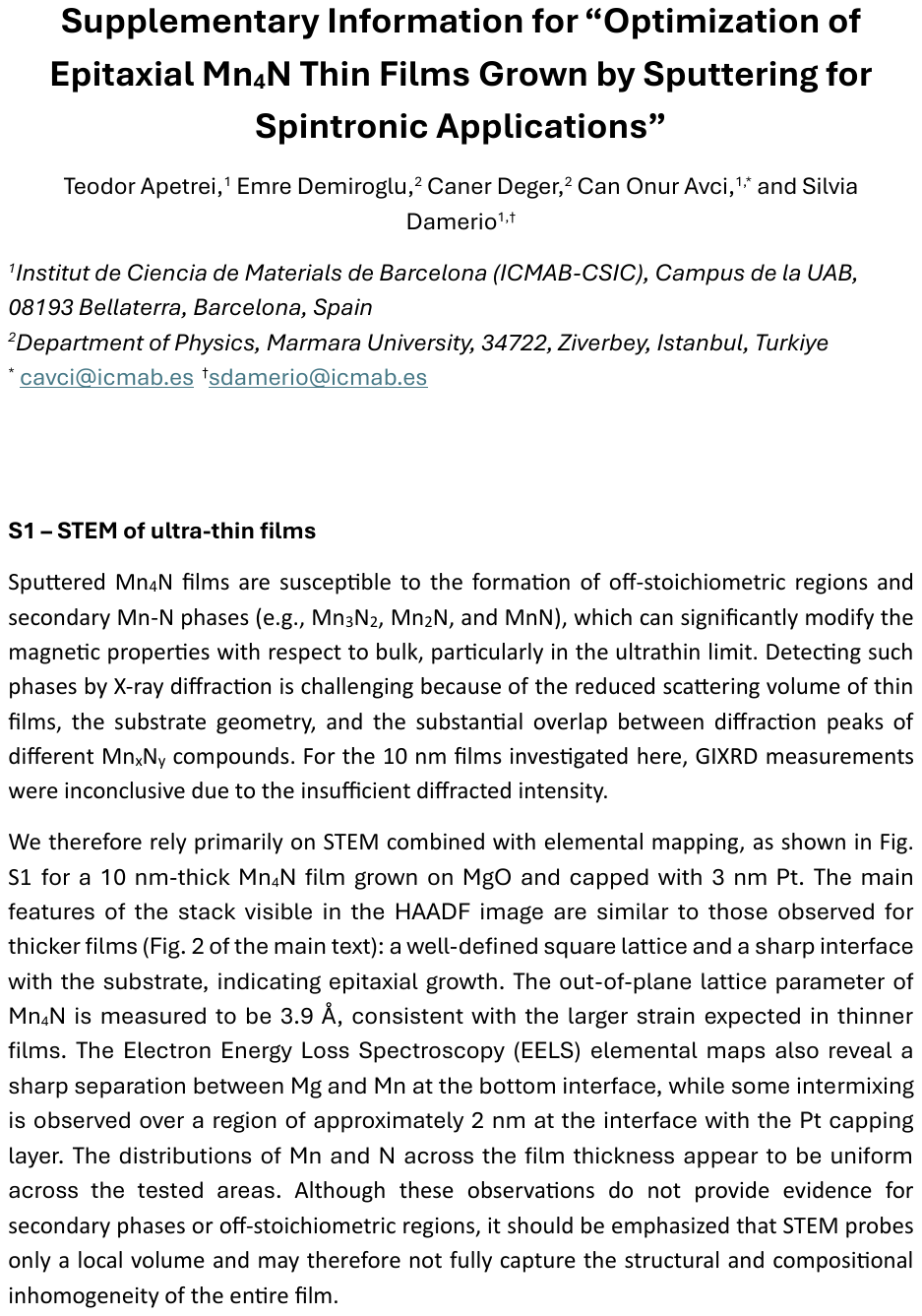}
\end{center}

\clearpage
\begin{center}
\includegraphics[page=2,width=\textwidth,height=\textheight,keepaspectratio]{Supplementary_Information.pdf}
\end{center}

\clearpage
\begin{center}
\includegraphics[page=3,width=\textwidth,height=\textheight,keepaspectratio]{Supplementary_Information.pdf}
\end{center}

\clearpage
\begin{center}
\includegraphics[page=4,width=\textwidth,height=\textheight,keepaspectratio]{Supplementary_Information.pdf}
\end{center}

\clearpage
\begin{center}
\includegraphics[page=5,width=\textwidth,height=\textheight,keepaspectratio]{Supplementary_Information.pdf}
\end{center}

\clearpage
\begin{center}
\includegraphics[page=6,width=\textwidth,height=\textheight,keepaspectratio]{Supplementary_Information.pdf}
\end{center}

\clearpage
\begin{center}
\includegraphics[page=7,width=\textwidth,height=\textheight,keepaspectratio]{Supplementary_Information.pdf}
\end{center}

\clearpage
\begin{center}
\includegraphics[page=8,width=\textwidth,height=\textheight,keepaspectratio]{Supplementary_Information.pdf}
\end{center}
\end{document}